\documentclass[runningheads]{llncs}
\usepackage[T1]{fontenc}
\usepackage{comment}
\usepackage{enumerate}
\usepackage{graphicx, color}                       
\usepackage[utf8]{inputenc}
\usepackage{paralist}
\usepackage[english]{babel}
\usepackage{url}
\usepackage{tabularx,ragged2e}                        
\usepackage{verbatim}
\usepackage{balance}                          
\usepackage{multirow}     
\usepackage{multicol}            
\usepackage{fancyvrb}       
\usepackage{xspace}
\usepackage{amsmath}
\usepackage{lipsum}
\usepackage{tcolorbox}
\usepackage{makecell}
\usepackage{listings}
\usepackage{todonotes}
\usepackage{colortbl}
\usepackage{longtable} 
\usepackage{enumitem}
\usepackage{textalpha}
\usepackage{supertabular,booktabs}
\usepackage{fontawesome}
\usepackage{subcaption}
\usepackage{hyphenat}
\usepackage{hyperref}
\usepackage{nicematrix}
\usepackage{booktabs}
\usepackage{array}
\usepackage{colortbl}
\usepackage{graphicx}
\usepackage{color}

\newcommand{\resquestion}[2]{ %
	\vspace{5pt} %
	\noindent\fcolorbox{black}{blue!05}{%
		\parbox{0.97\linewidth}{%
			\textbf{RQ$_{#1}$.} \emph{#2} %
		}%
	}%
	\vspace{5pt} %
}%

\newcommand{\ranswer}[2]{ %
	\vspace{5pt} %
	\noindent\fcolorbox{black}{blue!05}{%
		\parbox{0.97\linewidth}{%
			\textbf{\faChevronRight  \ RQ$_{#1}$.} #2 
		}%
	}%
	\vspace{5pt} %
}%

\begin{document}
\title{Rookie Mistakes: Measuring Software Quality in Student Projects to Guide Educational Enhancement}
\titlerunning{Software Quality in Student Projects to Guide Educational Enhancement}
%
\author{Marco De Luca\inst{1}\orcidID{0000-0002-9041-9088} \and
Sergio Di Martino\inst{1}\orcidID{0000-0002-1019-9004} \and
Sergio Di Meglio\inst{1}\orcidID{0009-0002-2224-4631} \and
Anna Rita Fasolino\inst{1}\orcidID{0000-0001-7116-019X} \and
Luigi Libero Lucio Starace\inst{1}\orcidID{0000-0001-7945-9014} \and Porfirio Tramontana\inst{1}\orcidID{0000-0003-3264-185X} }
\authorrunning{De Luca et al.}
%
\institute{University of Naples ``Federico II'', Naples, Italy
\email{\{marco.deluca2,fasolino sergio.dimartino, sergio.dimeglio, annarita.fasolino, luigiliberolucio.starace, ptramon\}@unina.it}
}
\maketitle              
\begin{abstract}
When teaching Programming and Software Engineering in Bachelor's Degree programs, the emphasis on creating functional software projects often overshadows the focus on software quality, a trend that aligns with ACM curricula recommendations. Software Engineering courses are typically introduced later in the curriculum, and can generally allocate only limited time to quality-related topics, leaving educators with the challenge of deciding which quality aspects to prioritize. In this decision, the literature offers limited guidance, as most existing studies focus on code written by novice students and small code units, making it unclear whether those findings extend to intermediate-level students with foundational object-oriented programming skills working on more complex software projects. To address this gap, we analyze 83 object-oriented team projects developed by 172 university students across 4 different editions of the Object-Oriented Programming course. We apply a static analysis pipeline used in prior research to assess software quality, combining \textsc{SonarQube} and \textsc{ArchUnit} to detect code smells and architectural anti-patterns. Our findings highlight recurring quality issues and offer concrete evidence of the challenges students face at this stage, providing valuable guidance for educators aiming to continuously improve Software Engineering curricula and promote quality-oriented development practices.

\end{abstract}
\section{Introduction}

In Bachelor's Degree programs, such as Computer Science and Computer Engineering, the focus of programming and software engineering education often leans more towards the ability to develop functional projects rather than emphasizing software quality. This aligns with the ACM Computer Science and Computer Engineering curriculum recommendations for undergraduate degrees \cite{clear2019computing}. According to these guidelines, software quality is typically only lightly addressed in three-year Bachelor's programs. In these programs, introductory CS1 courses concentrate primarily on programming skills, with basic concepts of software quality introduced later, primarily in Software Engineering courses.

However, even within Software Engineering courses, the time dedicated to software quality teaching is limited. Educators face the challenge of determining which aspects of software quality to prioritize within these constraints. Furthermore, there is a significant gap in the literature regarding research into the foundational software quality challenges specifically faced by intermediate-level students when developing software projects. This lack of insight leaves instructors of more advanced Software Engineering courses with little direction on how to prioritize quality aspects in their curricula.

\begin{figure}
    \centering
    \includegraphics[width=0.6\linewidth]{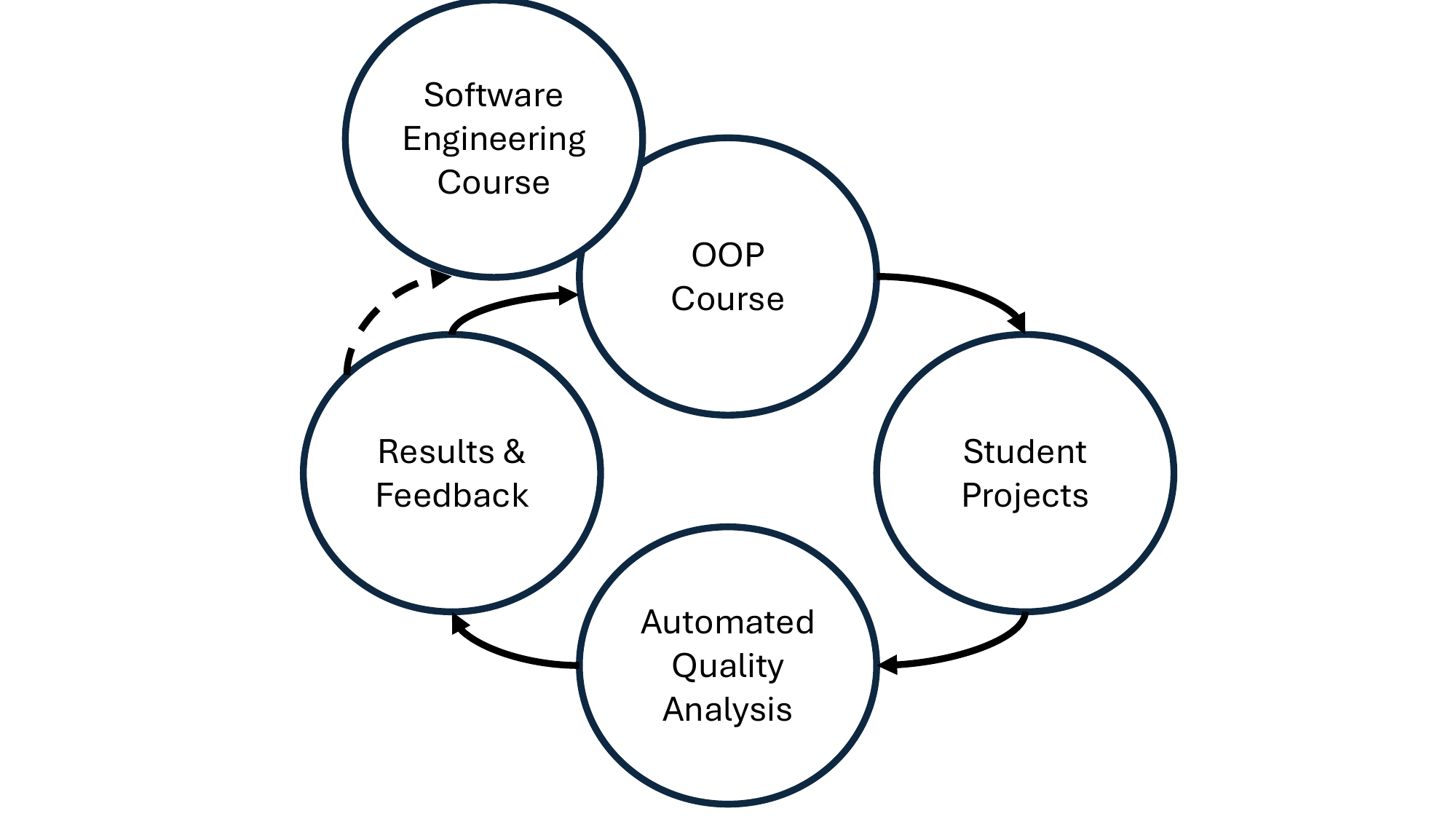}
    \caption{Cyclical evaluation process: the analysis of the quality of OOP students projects brings feedback useful to improve this course and also Software Engineering courses.}
    \label{fig:pipeline}
\end{figure}

In this study, we present an empirical investigation of software quality in a set of 83 projects developed by 172 Computer Science students involved in an Object-Oriented Programming (OOP) course taken in two consecutive academic years. Our study leverages a static code analysis pipeline validated in prior research \cite{de2024automatic}, applying state-of-the-art tools such as \textsc{SonarQube}\footnote{SonarQube, available at \href{https://www.sonarsource.com/products/sonarqube/}{https://www.sonarsource.com/products/sonarqube/}} and \textsc{ArchUnit}\footnote{ArchUnit, available at \href{https://www.archunit.org/}{https://www.archunit.org/}}  to detect code smells and architectural anti-patterns, two indicators of structural degradation and design compromise often referred to as social debt when accumulated in collaborative environments. 
By identifying recurring quality problems in student code, this work provides concrete evidence to support iterative course improvements aligned with real-world development standards \cite{yurkofsky2020research}, as shown in Figure \ref{fig:pipeline}. The insights gained from this analysis can guide educators in refining both introductory programming courses and more advanced Software Engineering modules, ultimately fostering a culture of quality awareness and reflective development practices in future software engineers.

The remainder of the paper is structured as follows. Section \ref{sec:related} presents background information on software quality and related work. Section \ref{sec:study} describes the study we conducted, including detailed information about the OOP courses and the topics covered, allowing readers to compare them with their own teaching contexts. Section \ref{sec:res} presents the results, highlighting the most relevant findings, which are then discussed in Section \ref{sec:discussion}, with a focus on their implications for educators. 
Last, Section \ref{threats} outlines the threats to the validity of our findings, while Section \ref{sec:conclusions} presents final remarks and directions for future work.

\section{Background and Related Works}\label{sec:related}

In the following, we introduce key concepts related to software quality, focusing on both code-level and architectural aspects. We also review existing tools used for quality assessment and summarize prior research on software quality issues in student-developed code.

\subsection{Software Quality}
\label{sec:backgroundAndRelatedWork}

Software quality captures how effectively a system satisfies its specified requirements and fulfills user expectations \cite{software.quality.def}. The ISO 25010 standard \cite{iso25010} provides a widely adopted model for assessing software quality, defining key attributes such as modularity, reusability, and testability \cite{di2023starting}. These quality attributes are primarily shaped by two fundamental aspects: the clarity and maintainability of the source code, and the soundness of the software’s architectural design. Both play a critical role in ensuring the system remains robust, adaptable, and easy to evolve over time. To ensure and maintain these properties, various Software Quality Management (SQM) practices are employed throughout the development process. These include code reviews, automated testing, static code analysis, and architecture conformance checks \cite{code_review,sqm1,sqm_testing}, often supported by tools such as \textsc{SonarQube}, \textsc{CheckStyle}, and similar analyzers. In recent years, SQM has also extended to the social dimension of software development. Community detection techniques have been used on software dependency graphs to identify architectural issues, and on developer collaboration networks to uncover coordination problems, knowledge silos, or uneven code ownership \cite{sqm_comm1,sqm_com2,sqm_comm4,sqm_comm3}. These analyses provide complementary insights to technical evaluations, enabling a more comprehensive understanding of software quality.

\subsubsection{Source Code Quality}

The notion of source code quality lacks a single, universally accepted definition and is often interpreted in various ways. One of the most well-known and widely used interpretations is the notion of \textit{Clean Code}, as introduced by Martin et al. \cite{martin2009clean}. According to this perspective, clean code is characterized by its readability, clarity of intent, ease of writing, and maintainability. It is designed not only to be understood by compilers, but also to be easily interpreted by other developers.

To support the development of such code, a variety of static analysis tools have been introduced and are commonly adopted in both industry and education. These tools are intended to automatically detect violations of clean code principles. A comparative study by Lenarduzzi et al.~\cite{LENARDUZZI2023111575} evaluated several popular tools, including \textsc{Better Code Hub}, \textsc{CheckStyle}, \textsc{Coverity Scan}, \textsc{FindBugs}, \textsc{PMD}, and \textsc{SonarQube}. Their findings highlighted \textsc{SonarQube} as the most effective tool for detecting a broad range of code quality issues.

\subsubsection{Software Architectural Design}
\label{sec:sad}

The architecture of a software system, defined as its structural organization into components, their relationships, and the rules governing their interactions, plays a critical role in determining overall system quality \cite{Aldrich.SoftArch,10741246}. A well-structured architecture contributes to building systems that are robust, scalable, and maintainable, while inadequate architectural decisions can lead to technical debt, reduced performance, and decreased reliability. To address recurring challenges in system design, software architectural patterns have been proposed as standardized, reusable solutions to common design problems \cite{taylor2010software}. These patterns serve as high-level design strategies that promote important architectural qualities such as modularity, reusability, and scalability \cite{book_sei,shaw1996software}. By guiding the organization of system components, they help ensure that software systems are easier to evolve and maintain over time. To verify that the implemented architecture adheres to the intended design, developers can perform Architecture Conformance Checks (ACC) \cite{Zak2020}. These checks can be automated using dedicated tools and libraries, such as \textsc{ArchUnit}, which allow developers to define and enforce architectural rules directly within the codebase.

\subsection{Software Quality Issues in Student Code}

Prior research has extensively examined the quality of code produced by university students, particularly those at the introductory level or with limited programming experience. Several studies (e.g., \cite{Jansen2017,Gomes17,Lu2019}) have explored the use of automated analysis tools to monitor and improve the quality of student code throughout a course, providing continuous feedback as a means of supporting the learning process. While these approaches have demonstrated a positive influence on student outcomes, their primary focus has been on measuring quality improvements over time, rather than uncovering the underlying challenges faced by novice programmers. Other studies have sought to identify recurring quality issues in student-written code by analyzing large sets of code submissions after completion. For instance, Keuning et al. \cite{CodeQualityIssue2017} examined over 2.6 million code snapshots created using the BlueJ environment to uncover frequent problems encountered by beginners. Similarly, Effenberger and Pelánek \cite{CodeQualityPythonShort2022} analyzed more than 114,000 Python solutions to small-scale programming tasks—typically no longer than 20 lines of code—to construct a taxonomy of common code defects among students in CS1-level courses. A more recent study by Sun et al. \cite{sun2020toward} focused on the performance of students in OOP courses, assessing both the quality of their code and the tests they wrote. However, this study also centered on relatively small assignments that did not involve substantial architectural design decisions. In a different educational context, Chren et al. \cite{Chren22} evaluated how a course explicitly focused on software quality impacted the code developed by students. Their evaluation considered 54 student projects using a combination of manual inspection and automated tools (e.g., \textsc{SonarQube} and \textsc{CheckStyle}), assessing multiple metrics such as code size, duplication, bugs, and code smells.

Despite the breadth of these studies, most are limited in two important ways. First, they typically examine simple, self-contained tasks, often solvable in just a few lines of code. This raises questions about whether their conclusions hold when students are tasked with more complex, OOP projects that integrate components such as graphical user interfaces (GUIs) and relational databases. Second, due to the simplicity of the programming problems under consideration, prior work has rarely addressed architectural quality issues, such as violations of design patterns or the emergence of architectural anti-patterns. To our knowledge, these limitations reduce the applicability of existing research to intermediate-level students working on structured, multi-component software systems, as found in typical OOP courses. In such contexts, the scope and architectural complexity of student projects differ significantly from those studied in most prior work.

A first exploratory study based on some projects from a single course was previously presented by the same authors \cite{de2024automatic} as a proof-of-concept of the idea of mining projects made by students of an OOP course for finding the presence of anti-patterns and code smells.

\section{Study Design}
\label{sec:study}

\subsection{Goal and Research Questions}

The goal of this study is to assess the software quality of Java projects developed by students attending an OOP course. These courses typically mark the students’ first exposure to the principles of designing and implementing software systems using the object-oriented paradigm. By examining the students’ code, we aim to identify recurring design flaws and code smells and better understand the foundational quality issues they face at this stage of their learning.

To this end, we conducted an empirical study involving two different editions of an OOP course taught by two of the authors as part of a Computer Science bachelor's degree program. We applied a static analysis pipeline grounded in prior research to investigate the presence of structural and design issues,  as illustrated in Figure \ref{fig:pipeline}.

This study addresses the following research questions:

\resquestion{1}{How common are architectural pattern violations found in students’ projects?}

\resquestion{2}{How common are code quality issues in students’ projects?}

The research questions aim to identify the most common shortcomings in student-developed projects, both from an architectural and a programming perspective. The findings are intended to provide actionable insights that can help educators improve both introductory OOP courses and more advanced courses, such as Software Engineering, by guiding them on which quality aspects to prioritize in their teaching.

\subsection{Object-Oriented Programming Course}

The objects of this study are the software projects developed by students as the final examination assignment for an OOP course.

We considered two editions of two OOP courses offered within the same Bachelor's degree program in Computer Science at the same university. These two courses share the same teaching materials, structure, and learning objectives, and are delivered during the same semester. Due to the large number of enrolled students, the cohort is divided into two groups based on students’ surnames. Each group is assigned to a different teacher, both of whom are coauthors of this study. To account for possible variation over time, we included projects from two consecutive academic years, namely 2021/2022 and 2022/2023. For clarity, we refer to the courses taught by the two teachers as Group 1 and Group 2 throughout the paper.

\begin{figure}
    \centering
    \includegraphics[width=.6\linewidth]{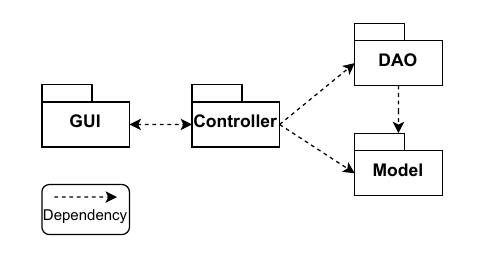}
    \caption{UML Package diagram representing the project architecture according to the proposed pattern.}
    \label{fig:OOPattern}
\end{figure}

The OOP courses included 48 hours of lectures, corresponding to 6 ECTS, in the standard European Credit Transfer and Accumulation System (ECTS) way of defining the academic characteristics of courses. The prerequisites for these courses include successful completion of prior CS1 programming courses included in the bachelor's degree program, which entail 72 and 24 lecture hours, respectively. In these courses, students learned the fundamentals of procedural programming using the C language.
Moreover, an additional prerequisite was completing the Database course, which is provided in the same year as the OOP course.
The OOP courses introduce students to the Java programming language, OOP principles, the UML modeling language, and basic software quality and architecture concepts. The official reference textbooks for the course are \textit{``Java How to Program"} \cite{librotesto_java}, and \textit{``UML Distilled: A Brief Guide to the Standard Object Modeling Language"} \cite{librotesto_uml}. The teachers did not require the use of a specific IDE to support development, but Eclipse was the most used by these classes of students.
As part of the course, students are also introduced to a reference architectural pattern designed for GUI-based applications. This pattern organizes the system into four main packages:  (1) a GUI package (2) a Controller package, (3) a Model package, (4) a Data Access Object (DAO) package, with, possibly, another package with utility classes. The proposed reference architecture is sketched in Figure \ref{fig:OOPattern}.

\color{black}

The GUI Package includes GUI classes representing the user entry point to the application. These classes should depend on Controller classes, implementing the application logic. 
Controller classes, on the other hand, depend on Model classes that are responsible for the transient storage of the information domain data and on DAO  classes for accessing the persistent data stored in the database. Controller classes are allowed to open GUI instances. 
The teachers recommended to follow this architectural pattern and to avoid the introduction of too much dependencies between packages. In particular, they recommended that the GUI should remain independent from data management components. Although this pattern may not be representative of other patterns adopted in complex software systems (e.g. MVC), they adopted it as a didactic introduction to the principle of separation of concerns and to avoid an excessive coupling between the application's components.

To pass the course, students have to submit a final assignment consisting of developing a GUI-based desktop application, representing a typical information system, relying on a relational database for data persistence. In this final assignment, students have to self-organize in teams of two or three students and use the Java programming language and the reference architecture presented during the course. 
The project requirements are assigned by the teacher to each team. In each course edition, three different sets of project requirements, each with comparable complexity, were defined and randomly assigned to each team.
The teachers graded the submitted projects based on their completeness and correctness w.r.t. the assigned requirements, and overall software quality. 
Each student was also subjected to a final oral exam during which the project was discussed. This final moment of evaluation was useful to obtain feedback from the students regarding their choices and the mistakes they had made.

In these editions of the Object Orientation course the teachers did not explicitly present the concept of code smell, nor did they present analysis tools such as ArchUnit and SonarQube so the primary goal of the experiments presented in this paper is to show the issues that students most frequently introduce without an explicit education on software quality assessment. 

\begin{table}
\scriptsize
\centering
\caption{Characteristics of the considered projects}
\label{tab:statistics}
\begin{tabular}{llccrrrr}
\toprule
\multirow{2}{*}{\textbf{Course}} & \multirow{2}{*}{\textbf{Year}} & \multirow{2}{*}{\makecell{\textbf{Num.}\\\textbf{Proj.}}} & \multirow{2}{*}{\makecell{\textbf{Num.}\\\textbf{Stud.}}} & \multicolumn{2}{c}{\bfseries Num. of classes} & \multicolumn{2}{c}{\bfseries NCLOC} \\
\cmidrule(lr){5-6} \cmidrule(lr) {7-8}
 & & & & \bfseries Mean & \bfseries St.d. & \bfseries Mean & \bfseries St.d. \\ 
  \midrule
\multirow{2}{*}{\textbf{Gr. 1}} 
& 21/22 & 23 & 46 & 51 & 26 & 5797 & 3099 \\ 
& 22/23 & 20 & 47 & 40 & 24 & 3887 & 2610 \\ 
\midrule
\multirow{2}{*}{\textbf{Gr. 2}} 
& 21/22 & 26 & 50 & 41 & 13 & 4647 & 1616 \\ 
& 22/23 & 14 & 29 & 36 & 11 & 3011 & 1171 \\ 
   \bottomrule
\end{tabular}

\end{table}

\begin{figure}[h!]
\centering
\includegraphics[width=0.4\linewidth]{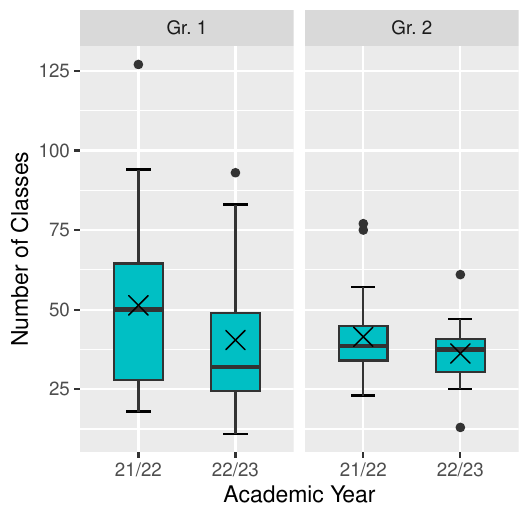}
\includegraphics[width=0.4\linewidth]{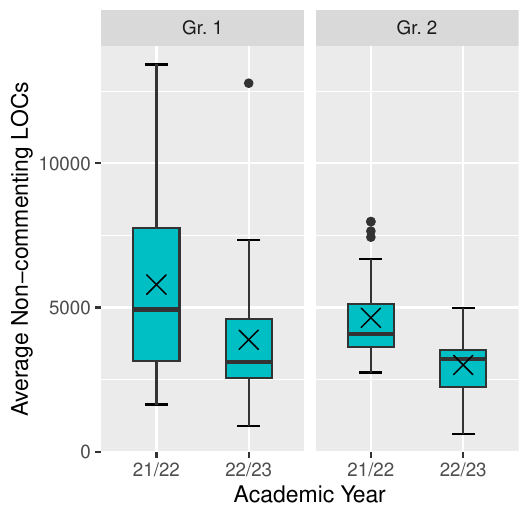}
\caption{Project Statistics. From left to right, boxplots represent the distribution of the number of classes and the number of non-commenting lines of code per project. The cross, in each boxplot, represents the average value.}
\label{fig:CourseStatistics}
\end{figure}

In total, as objects of our study, we collected $83$ projects developed by $172$ students.
Table \ref{tab:statistics} and Figure \ref{fig:CourseStatistics} summarize, for each course edition, the number of considered projects, the number of involved students, and some statistics about project size (mean number of classes and non-commenting lines of code - NCLOC, along with the standard deviation).
Note that these projects are remarkably more complex than the simple assignments employed in related works investigating software quality in student-developed code, with an average number of classes per project ranging from 36 to 51, depending on the edition of the course, and 3,000 or more NCLOC on average.

\subsection{Automated Quality Analysis}

To evaluate the student submissions, we adopted an automated quality analysis pipeline inspired by the approach proposed in \cite{de2024automatic}. The pipeline performs two distinct but complementary analyses: \textit{Architecture Conformity Check} and \textit{Static Code Quality Analysis}. The Architecture Conformity Check leverages \textsc{ArchUnit}, a Java library that allows developers to encode architectural rules as JUnit tests. A suite of \textsc{ArchUnit} test cases was created by two of the authors to verify compliance with the architectural patterns taught during the course. These tests systematically check for disallowed dependencies between packages and flag any violations. For each type of violation, we recorded the number of student projects where it occurred at least once, and we qualitatively examined frequent violations to understand their underlying causes. For the Code Quality Analysis, we relied on \textsc{SonarQube}, a widely used static analysis platform. We configured a dedicated \textsc{SonarQube} instance to analyze the codebases against the full set of 677 default Java rules. \textsc{SonarQube} classifies issues into Bugs (reliability), Vulnerabilities (security), and Code Smells (maintainability), and assigns a severity level ranging from Info to Blocker. For each rule, we recorded the number of student projects in which at least one violation was detected.

\section{Results}
\label{sec:res}

\textit{\textbf{RQ$_1$: How common are architectural pattern violations found in students’ projects?}} 
\newline
Table \ref{tab:ProjectDependencies} reports the number and percentage of projects from each considered group that presented at least one occurrence of a disallowed type of coupling (respectively from GUI to Model, from Model to GUI, from GUI to DAO, from DAO to GUI, from Model to Controller, from DAO to Controller and from Model to DAO).

\begin{table}[htbp]
  \centering
  \tiny
    \caption{  Overview of package dependencies across projects. Disallowed dependencies are highlighted in red. Legend: G = GUI, C = Controller, M = Model, D = DAO.}
    \label{tab:ProjectDependencies}    
    \begin{tabular}{ccrrrrrrrrrrrrr}
    \toprule
    \multicolumn{3}{c}{}  & \multicolumn{12}{c}{\textbf{Couplings}} \\
    \cmidrule{4-15}
    \multicolumn{1}{l}{\textbf{Gr}} & \multicolumn{1}{l}{\textbf{Year}} & \multicolumn{1}{l}{\textbf{\#Proj}} & \multicolumn{1}{l}{\textcolor[rgb]{ 1,  0,  0}{\textbf{M$\rightarrow$G}}} & \multicolumn{1}{l}{\textbf{C$\rightarrow$G}} & \multicolumn{1}{l}{\textcolor[rgb]{ 1,  0,  0}{\textbf{D$\rightarrow$G}}} & \multicolumn{1}{l}{\textcolor[rgb]{ 1,  0,  0}{\textbf{G$\rightarrow$M}}} & \multicolumn{1}{l}{\textbf{C$\rightarrow$M}} & \multicolumn{1}{l}{\textbf{D$\rightarrow$M}} & \multicolumn{1}{l}{\textbf{G$\rightarrow$C}} & \multicolumn{1}{l}{\textcolor[rgb]{ 1,  0,  0}{\textbf{M$\rightarrow$C}}} & \multicolumn{1}{l}{\textcolor[rgb]{ 1,  0,  0}{\textbf{D$\rightarrow$C}}} & \multicolumn{1}{l}{\textcolor[rgb]{ 1,  0,  0}{\textbf{G$\rightarrow$D}}} & \multicolumn{1}{l}{\textcolor[rgb]{ 1,  0,  0}{\textbf{M$\rightarrow$D}}} & \multicolumn{1}{l}{\textbf{C$\rightarrow$D}} \\
    \midrule
    \multicolumn{1}{r}{1} & \multicolumn{1}{l}{21/22} & 23    & \textcolor[rgb]{ 1,  0,  0}{9\%} & 100\% & \textcolor[rgb]{ 1,  0,  0}{0\%} & \textcolor[rgb]{ 1,  0,  0}{65\%} & 91\%  & 91\%  & 96\%  & \textcolor[rgb]{ 1,  0,  0}{13\%} & \textcolor[rgb]{ 1,  0,  0}{35\%} & \textcolor[rgb]{ 1,  0,  0}{22\%} & \textcolor[rgb]{ 1,  0,  0}{4\%} & 96\% \\
    \multicolumn{1}{r}{1} & \multicolumn{1}{l}{22/23} & 20    & \textcolor[rgb]{ 1,  0,  0}{0\%} & 95\%  & \textcolor[rgb]{ 1,  0,  0}{15\%} & \textcolor[rgb]{ 1,  0,  0}{75\%} & 90\%  & 85\%  & 95\%  & \textcolor[rgb]{ 1,  0,  0}{10\%} & \textcolor[rgb]{ 1,  0,  0}{45\%} & \textcolor[rgb]{ 1,  0,  0}{35\%} & \textcolor[rgb]{ 1,  0,  0}{10\%} & 100\% \\
      \midrule
    \multicolumn{1}{r}{2} & \multicolumn{1}{l}{21/22} & 26    & \textcolor[rgb]{ 1,  0,  0}{0\%} & 15\%  & \textcolor[rgb]{ 1,  0,  0}{4\%} & \textcolor[rgb]{ 1,  0,  0}{88\%} & 96\%  & 96\%  & 100\% & \textcolor[rgb]{ 1,  0,  0}{4\%} & \textcolor[rgb]{ 1,  0,  0}{4\%} & \textcolor[rgb]{ 1,  0,  0}{31\%} & \textcolor[rgb]{ 1,  0,  0}{0\%} & 88\% \\
    \multicolumn{1}{r}{2} & \multicolumn{1}{l}{22/23} & 21    & \textcolor[rgb]{ 1,  0,  0}{0\%} & 50\%  & \textcolor[rgb]{ 1,  0,  0}{7\%} & \textcolor[rgb]{ 1,  0,  0}{64\%} & 100\% & 57\%  & 93\%  & \textcolor[rgb]{ 1,  0,  0}{0\%} & \textcolor[rgb]{ 1,  0,  0}{0\%} & \textcolor[rgb]{ 1,  0,  0}{7\%} & \textcolor[rgb]{ 1,  0,  0}{0\%} & 93\% \\
    \midrule
    \multicolumn{2}{c}{\textbf{Overall}} & \textbf{83} & \textcolor[rgb]{ 1,  0,  0}{\textbf{2\%}} & \textbf{64\%} & \textcolor[rgb]{ 1,  0,  0}{\textbf{6\%}} & \textcolor[rgb]{ 1,  0,  0}{\textbf{75\%}} & \textbf{94\%} & \textbf{86\%} & \textbf{96\%} & \textcolor[rgb]{ 1,  0,  0}{\textbf{7\%}} & \textcolor[rgb]{ 1,  0,  0}{\textbf{22\%}} & \textcolor[rgb]{ 1,  0,  0}{\textbf{25\%}} & \textcolor[rgb]{ 1,  0,  0}{\textbf{4\%}} & \textbf{94\%} \\
    \bottomrule
    \end{tabular}%
  \label{tab:addlabel}%
\end{table}%

The most common violations of the expected architectural pattern were due to the existence of dependencies from the GUI package to the Model package (G$\rightarrow$M) (75\%). 
The direct dependencies from GUI to Model (G$\rightarrow$M) were not allowed because in these courses the teachers recommended that Model classes and objects should be managed only by Controllers and not directly by GUI classes to make the GUI implementation independent from the information modelling. We observed this architectural issue very often, in all the courses (in 75\% of projects). 
\textit{These dependencies usually represented a shortcut adopted by students to avoid the need to serialize data from model classes in data structures for the GUI.} On the other hand, we observed that the inverse coupling (M$\rightarrow$G), due to model classes that request data from the GUI, was correctly avoided in almost all the projects, except 5 of them (2\%).

The direct interactions between classes of the GUI package and DAO classes (G$\rightarrow$D and D$\rightarrow$G) were sometimes observed: on average, 25\% of projects included calls from GUI to DAO, whereas only 6\% of projects included calls from DAO to GUI. These interactions are violations causing the GUI to depend on the database and on its data representation, or vice versa. \textit{Also these interactions were introduced by students as shortcuts but they caused unneeded couplings betweeen packages that should remain unlinked between them.  }

The unallowed interaction between DAO and Controller (D$\rightarrow$C) was observed in a significant set of projects, almost all belonging to Group 1 (22\% on average).
The DAO object should only return data to the Controller, without other calls. \textit{We observed that some student teams included part of the business logic into the methods of the DAO classes, maybe confusing those methods with database triggers}.

We observed a few cases (only in 7\% of projects) in which classes of the Model package interacted with the Controller (M$\rightarrow$C). Also in that case, these interactions corresponded to methods of the Model that carried out part of the business logic that should be an exclusive responsibility of Controllers.

Finally, we observed some cases in which objects of the Model package directly queried the database by calling items from the DAO package. Although this interaction is allowed in some simple architectural patterns (i.e. Boundary-Control-Entity-Database, BCED), it represents a disallowed interaction in the suggested pattern.
These interactions were observed in few projects (only 3 projects from the Group 1, consisting in 4\% of the total number of projects).

\ranswer{1}{Overall, the students' projects presented 7 different types of violations of the prescribed architectural pattern. The frequencies of these violations ranged between 2\% (Model$\rightarrow$GUI) and 75\% (GUI$\rightarrow$Model). Most of the violations were essentially due to (1) bad practices (e.g. shortcuts) adopted by the students to avoid serialization and deserialization operations for passing data between classes of different packages, and (2) inadequate understanding of the principles for correctly assigning responsibilities to classes and for sharing them between specialized packages.}

\textit{\textbf{RQ$_2$: How common are code quality issues in students’ projects?}} 
\newline
Table \ref{tab:totalIssues} reports, for each course group and edition, the number of analysed projects and the average number of different types of code issues detected in all the projects belonging to that specific group and edition.

\begin{table}[]
\scriptsize
\centering
\caption{Overall number of different code issues typologies observed in student's projects}\label{tab:totalIssues}
\begin{tabular}{llcc}
\toprule
\textbf{Course}                 & \textbf{Year}  & \makecell{\bf Number of \\ \bf Projects} & \makecell{\textbf{Average Number} \\\textbf{Issue types}} \\ \midrule
\multirow{2}{*}{Gr. 1} & 21/22 & 23  & 33.1  \\
                       & 22/23 & 20  & 30.8   \\ \midrule
\multirow{2}{*}{Gr. 2} & 21/22 & 26  & 36.8    \\
                       & 22/23 & 14  & 27.9     \\
\bottomrule
\end{tabular}
\end{table}
The average number of typologies of detected issues varied from 33.1 and 36.8, measured in blended course editions (21/22) and 30.8 and 27.9, measured in the 22/23 editions of the courses.
A possible cause for these slight reductions in the number of detected code smells per project might be due to the increased effectiveness of teaching in person.
In addition, in the second editions of the courses, both the teachers presented in more detail the ``Good Programming Practice" and the ``Common Programming Error" boxes present in the textbook ``Java: How to Program" \cite{librotesto_java} in order to better focus on these aspects. \color{black}

Complete tables, including the entire list of smell typologies observed in all the student projects, are included in the anonymized replication package \cite{4openAnonymousGithub}.
To have a summary view of the most common issues in the selected student projects, we limited the scope of our analysis to the ones occurring in the majority (more than 50\%) of the projects. We have filtered out from this set some issues that were explicitly related to topics outside the scope of the course (e.g., the lack of use of lambda expressions, which are introduced in advanced programming courses). As a result of this filtering step, 17 types of issues were selected. With respect to their severity, one of them is classified as Blocker, 3 as Critical, 7 as Major, and the remaining 6 as Minor. Only one of them is considered a Bug, whereas all the other ones are classified as Code Smells. Moreover, no issues labelled with the Vulnerability tag were found in the majority of projects. 16 out of 17 issues affect the maintainability of the project, whereas the remaining one (i.e., the Bug) affects its reliability.

Table \ref{tab:smellDetails} reports the list of the 17 issue types found in the majority of the examined student projects. For each issue type, the table reports its description as provided by \textsc{SonarQube}. In the other columns, we reported the classification of the issue (Bug or Code Smell) and its severity (Blocker, Critical, Major, or Minor). The last columns of the table report the number (and percentage) of students' projects of each group in which the issue type was detected.

\begin{table*}[]
\setlength{\tabcolsep}{.5em}
\centering
\scriptsize
\caption{List of the most common code issues found in student projects}
\rotatebox{90}{
\begin{tabular}{lccccccc}
\toprule
\multirow{2}{*}{\bf Sonar Rule Description} & \multirow{2}{*}{\makecell{\bf Issue\\\bf Type}}  & \multirow{2}{*}{\bf Severity} & \multicolumn{2}{c}{\bf Group 1} & \multicolumn{2}{c}{\bf Group 2} & \multirow{2}{*}{\makecell{\bf Overall\\\bf(\%)}}\\
\cmidrule(lr){4-5}\cmidrule(lr){6-7}
&   &     & 21/22        & 22/23        & 21/22        & 22/23        &        \\
\midrule
Resources should be closed                                                              & BUG         & BLOCKER  & 16 (70\%)    & 17(77\%)     & 22 (85\%)    & 10 (71\%)    & 76\%    \\ \midrule
String literals should not be duplicated                                                & SMELL       
& CRITICAL & 23 (100\%)   & 21 (95\%)    & 26 (100\%)   & 12 (86\%)    & 96\%    \\
Cognitive Complexity of methods should not be too high                                  & SMELL       & CRITICAL & 18 (78\%)    & 7 (32\%)     & 25 (96\%)    & 10 (71\%)    & 71\%    \\
Methods should not be empty                                                             & SMELL       & CRITICAL & 9 (39\%)     & 15 (68\%)    & 14 (54\%)    & 6 (43\%)     & 52\%    \\ \midrule
Standard outputs should not be used directly to log anything                            & SMELL       & MAJOR    & 18 (78\%)    & 18 (82\%)    & 24 (92\%)    & 13 (93\%)    & 86\%    \\
Unused assignments should be removed                                                    & SMELL       & MAJOR    & 17 (74\%)    & 16 (73\%)    & 21 (81\%)    & 6 (43\%)     & 71\%    \\
Methods should not have too many parameters                                             & SMELL       & MAJOR    & 12 (52\%)    & 15 (68\%)    & 17 (65\%)    & 11 (79\%)    & 65\%    \\
Sections of code should not be commented out                                            & SMELL       & MAJOR    & 15 (65\%)    & 12 (55\%)    & 17 (65\%)    & 10 (71\%)    & 64\%    \\
Unused "private" fields should be removed                                               & SMELL       & MAJOR    & 15 (65\%)    & 12 (55\%)    & 17 (65\%)    & 9 (64\%)     & 62\%    \\
Local variables should not shadow class fields                                          & SMELL       & MAJOR    & 12 (52\%)    & 14 (64\%)    & 17 (65\%)    & 5 (36\%)     & 56\%    \\
Branches in a conditional structure should not have the same implementation             & SMELL       & MAJOR    & 6 (26\%)     & 2 (9\%)      & 23 (88\%)    & 12 (86\%)    & 51\%    \\ \midrule
Variables and method parameters should comply with naming conventions             & SMELL       & MINOR    & 21 (91\%)    & 22 (100\%)   & 25 (96\%)    & 12 (86\%)    & 94\%    \\
Field names should comply with a naming convention                                      & SMELL       & MINOR    & 19 (83\%)    & 20 (91\%)    & 23 (88\%)    & 9 (64\%)      & 84\%    \\
Unnecessary imports should be removed                                                   & SMELL       & MINOR    & 17 (74\%)    & 8(36\%)      & 25 (96\%)    & 14 (100\%)   & 75\%    \\
Method names should comply with a naming convention                                     & SMELL       & MINOR    & 14 (61\%)    & 19 (86\%)    & 18 (69\%)    & 5 (36\%)      & 66\%    \\
Unused local variables should be removed                                                & SMELL       & MINOR    & 18 (78\%)    & 15  (68\%)   & 16 (62\%)    & 6 (43\%)      & 65\%    \\
Class variable fields should not have public accessibility                              & SMELL       & MINOR    & 6  (26\%)    & 15  (68\%)   & 16 (62\%)    & 9 (64\%)      & 54\%   \\
\bottomrule
\end{tabular}
}
\label{tab:smellDetails}
\end{table*}

The Blocker category issue is concerned with Resources that were not properly closed. It was found in 76\% of projects. The issues were generally due to student inexperience: leaving open a stream or connection to an external resource can cause concurrency problems or memory leaks, but these topics are usually explained in third year courses.

The three Critical issues found in the majority of student projects are classified as Code Smells and are respectively ``\textit{String literals should not be duplicated}'', ``\textit{Cognitive Complexity of methods should not be too high}'', and ``\textit{Methods should not be empty}', respectively found in 96\%, 71\% and 52\% of projects. All of them negatively affected the maintainability of the projects. In particular, the first one may cause inconsistencies when string literals have to be modified (e.g. translated into another language), while the second one may make hard the comprehension of methods behaviour. The third one may be a symptom of project incompleteness.

The 7 most common issues categorized by \textsc{SonarQube} as Major issues correspond to other Code Smells negatively affecting the understandability and modifiability of the source code. For example, the use of \texttt{System.out} for logging purposes (found in 86\% of the projects) is not recommended because log outputs may mix with standard outputs and error outputs. 
Other frequent issues are related to unused assignments (71\%) or private fields (62\%), methods with too many parameters (65\%), old sections of code commented out (64\%), local variables with the same name of class fields (56\%), branches of the same conditional structure with the same implementation (51\%). All these bad practices appeared to be due to programmers' lack of attention to the quality of their code. Similarly, the 6 most common Minor issues corresponded to Code Smells. Many of them were due to a lack of coherence in the use of naming conventions about local variables, method parameters, and field and method names (e.g., starting with a capital letter). In addition, in 75\% of the projects, unused imports were found, and in 54\% of the projects, fields with public accessibility were used instead of private fields with possible public getter or setter methods. All these smells denoted a lack of knowledge about naming conventions and best practices that were explicitly presented during lectures.

It is worth noting that, even though not appearing in the majority of the projects, we observed that a security vulnerability was quite common. This vulnerability (classified as Blocker by \textsc{SonarQube}) was the hard-coding of database credentials in the source code of the projects (found in 39\% of projects). This issue is also due to student inexperience (most of the students developed the code locally, thus they did not consider the risks in sharing the source code, including private credentials).

\ranswer{2}{The analysis of students' projects revealed common code issues, mainly related to poor resource handling (unclosed resources in 76\% of projects) and programming bad practices. Frequent problems included duplicated string literals (96\%), excessive cognitive complexity (71\%), and empty methods (52\%). Naming convention violations were also widespread (up to 94\%), along with issues like unused assignments (71\%) and inappropriate public field accessibility (54\%).}

\section{Discussion and Implications}

\label{sec:discussion}

Building on the insights from our study, the following discusses the implications of these findings and suggests strategies to mitigate the issues observed, with the goal of improving both the courses and the quality of students' software engineering skills.

Regarding architectural violations, many were the result of students' limited experience, especially as they tackled their first substantial software projects. At this early stage, students often lack an understanding of how shortcuts or suboptimal decisions can negatively affect the long-term structure of a project. To address this, educators should place greater emphasis on explaining the \textit{why} behind architectural decisions, rather than just the \textit{what}. By providing clear examples and counterexamples that illustrate the consequences of poor architectural choices, educators can significantly help students make more informed and effective design decisions \cite{Ran2021,Zak2020}.
Most of the code smells observed in students' projects were linked to fundamental issues, such as unused code (e.g., redundant imports, fields, method parameters, and empty methods) and improper naming conventions for variables, fields, methods, and classes. These issues are typically the first encountered by students and can be effectively addressed with a stronger focus on clean code practices. At the introductory level, particularly in Object-Oriented Programming (OOP) courses, it would be beneficial to emphasize the most common and straightforward code smells, such as poor naming conventions, redundant variables, and unnecessary methods. Educators should encourage students to prioritize these foundational practices, as they directly improve code maintainability and readability. Teachers should stress the importance of writing clean, concise, and understandable code, illustrating how small adjustments can significantly enhance software quality.

As students progress to more advanced courses, such as Software Engineering, the curriculum should evolve to address a broader range of code smells, including more subtle and intricate issues that arise in larger-scale software projects. In these advanced courses, educators can introduce static analysis tools and more sophisticated techniques for identifying deeper code quality issues. One such tool, \textsc{SonarQube}, offers automated code quality checks and can help students integrate these tools into their development workflows.

Many code smells also stem from improper resource management, particularly with files, databases, and streams. This issue may be further compounded by students’ prior experience with C, where resource management requires explicit handling of file streams and memory allocation. Having learned to manually close resources in C, students may mistakenly believe that Java, with its automatic garbage collector, eliminates the need for explicit resource management.

This misunderstanding can lead to resource leaks when students neglect to close resources like file streams, resulting in inefficient resource usage and potential application crashes. To address this, educators should clarify the differences between C and Java in terms of resource management, emphasizing the importance of explicit resource handling in Java despite the presence of garbage collection. Introducing concepts such as the try-with-resources statement in Java can help students understand the necessity of managing resources properly.

\section{Threats to Validity} 
\label{threats}

Several factors may have influenced the conclusions drawn in this study. Below, we discuss the main limitations and mitigation strategies applied.

\begin{description}[leftmargin=0.3cm] 

\item[\textbf{Threats to Internal Validity.}] While this study investigates software quality in student projects, the results may be influenced by the detection capabilities of the tools employed, namely \textsc{SonarQube} and \textsc{ArchUnit}. Although these tools are the most widely adopted in the industry for static code analysis and are considered the state of the practice, they still have limitations in detecting certain types of code smells and architectural anti-patterns. We have attempted to mitigate this limitation by ensuring that \textsc{SonarQube} was configured according to best practices and that \textsc{ArchUnit} was specifically tuned and validated to detect violations in a way that closely aligns with industry standards. 

\item[\textbf{Threats to External Validity.}] The study's findings are based on projects developed by Computer Science students in two different OOP courses at a single university. As a result, the generalizability of the results to other contexts, such as different educational institutions, programming languages, or course structures, may be limited. To mitigate this, we have provided extensive details about the courses, including the curriculum, student demographics, and project scope. This allows other researchers to compare their own courses with ours and assess the extent to which our findings may apply to other settings. Additionally, while our study focused on detecting code smells and architectural anti-patterns, we acknowledge that these may not encompass all aspects of software quality.

\end{description}

\section{Conclusions and Future Work}
\label{sec:conclusions}

This study investigates the prevalence of architectural and code smells in Java projects developed by students in OOP courses. By analyzing 83 student projects, we identified recurring architectural violations and code quality issues, revealing the most common bad practices and their causes. Our findings suggest that both early programming courses (CS1) and advanced software engineering courses must address these issues to foster better design and coding practices. The results also highlight the importance of focusing on clean code principles and architectural decisions throughout the curriculum to improve software quality awareness.

Future work will focus on: (1) replicating this study in similar courses across different universities or degree programs (e.g., Software Engineering) to broaden the applicability of our findings; (2) extending the study to other Object Orientation classes that will be instead previously instructed about code smells, architectural anti-patterns and their automatic assessment, to evaluate the possible improvements of the quality of the projects;
(3) extending the evaluation to more complex student projects that integrate both front- and back-end components \cite{e2eloader-icst25,e2egit}, allowing a deeper investigation of architectural decisions, division of responsibilities, and the practical application of design principles in full-stack development scenarios. This will help assess the effectiveness of these improvements in fostering better design practices and enhancing the overall quality of student projects.

\begin{credits}
\subsubsection{\ackname} This work has been partially supported by the Italian PNRR MUR project PE0000013-FAIR and by the project GATT (GAmification in Testing Teaching), funded by the University of Naples Federico II Research Funding Program (FRA).
\end{credits}

\bibliographystyle{splncs04}
\bibliography{biblioStudentQuality}

\end{document}